\documentstyle[12pt]{article}

\newcommand\beq{\begin{equation}}
\newcommand\eeq{\end{equation}}
\def\beqa{\begin{eqnarray}}
\def\eeqa{\end{eqnarray}}

\def\obar{\overline}

\def\P{{\Psi}}
\def\C{{\cal C}}
\def\sH{{\cal H}}
\def\l{{\langle}}
\def\r{{\rangle}}

\begin{document}

\title{\vspace{-1cm}
Thawing the Frozen Formalism:\\ The Difference Between Observables\\
and What We Observe\\
\vspace{.2cm}
{\large\sl Essay for the Festschrift of Dieter Brill} }{}{}

\author{
Arlen Anderson\thanks{arley@physics.mcgill.ca}\\
Blackett Laboratory\\ Imperial College\\
Prince Consort Rd.\\ London SW7 2BZ England. }

\date{Nov. 23, 1992}

\maketitle
\vspace{-11.8cm}
\hfill Imperial/TP/92-93/09

\hfill gr-qc/9211028
\vspace{10cm}

\begin{abstract}
In a parametrized and constrained Hamiltonian system, an observable is an
operator which commutes with all (first-class) constraints, including the
super-Hamiltonian. The problem of the frozen formalism is to explain how
dynamics is possible when all observables are constants of the motion. An
explicit model of a measurement-interaction in a parametrized Hamiltonian
system is used to elucidate the relationship between three definitions of
observables---as something one observes, as self-adjoint operators, and as
operators which commute with all of the constraints. There is no
inconsistency in the frozen formalism when the measurement process is
properly understood. The projection operator description of measurement is
criticized as an over-idealization which treats measurement as
instantaneous and non-destructive. A more careful description of
measurement necessarily involves interactions of non-vanishing duration.
This is a first step towards a more even-handed treatment of space and time
in quantum mechanics.
\end{abstract}
\newpage



There is a special talent in being able to ask simple questions whose
answers reach deeply into our understanding of physics. Dieter is one of
the people with this talent, and many was the time when I thought the
answer to one of his questions was nearly at hand, only to lose it on
meeting an unexpected conceptual pitfall. Each time, I had come to realize
that if only I could answer the question, there were several interlocking
issues I would understand more clearly. In this essay, I will address such
a question posed to me by others sharing Dieter's talent:
\begin{quote}
What is the difference between an observable and what we observe?
\end{quote}

This question arises in the context of parametrized Hamiltonian systems, of
which canonical quantum gravity is perhaps the most famous example. It is
posed to resolve the following paradox: For constrained Hamiltonian
systems, an observable is defined as an operator which commutes (weakly)
with all of the (first-class) constraints. In the parametrized canonical
formalism, the super-Hamiltonian $\sH$ describing the evolution of states
is itself a constraint. Thus, all observables must commute with the
super-Hamiltonian, and so they are all constants of motion. Where then are
the dynamics that we see, if not in the observables?
This is the problem of the {\it frozen formalism}\cite{Fro,Rov,Kuc}.

In the context of quantum gravity, the problem of the frozen formalism is
closely linked with the problem of interpreting the wavefunction of
the universe and the problem of time.  Two proposed solutions
to the problem of time---Rovelli's evolving constants of the
motion\cite{Rov} and the conditional probability interpretation of Page
and Wootters\cite{PaW}---intimately involve observables which commute with
the super-Hamiltonian, and each claims to recover dynamics.  These
proposals have been strongly criticized by Kuchar\cite{Kuc}, who notes
that there is a problem with the frozen formalism even for the parametrized
Newtonian particle.

In this essay, I shall not address the problem of time but will focus on
the simpler case of the Newtonian particle. My intention is to reconcile
the different conceptions, mathematical and physical, that we have of
observables. This will involve a recitation of measurement theory to
establish the connection between the physical and the mathematical.
Essential features of both the Rovelli and the Page-Wootters approaches
will appear in my discussion as aspects of a careful understanding of
observables and how we use them.

There is a general consensus that to discuss the wavefunction of the
universe one must adopt a post-Everett interpretation of quantum theory in
which the observer is treated as part of the full quantum system. I shall
take this position for parametrized Hamiltonian systems as well. Insistence
that the measurement process must be explicitly modelled will lead to a
sharp criticism of the conventional description of measurement in terms of
projection operators. A simple model measurement, related to one originally
discussed by von Neumann\cite{vNe}, will clarify the role of observables in
the description of measurements. No incompatibility between dynamics and
observables which are constants of the motion will be found. With further
work, I believe that my discussion can be extended to answer some of the
criticisms of Kuchar of the Rovelli and the Page-Wootters proposals on the
problem of time.

Before beginning the analysis of observables, return to the formulation of
the problem of the frozen formalism. To be assured the problem doesn't lie
in the assumptions, consider each of the hypotheses leading up to it. The
stated definition of an observable is a sensible one as the following
argument shows. In a constrained Hamiltonian system, the set of
(first-class) constraints $\{ \C_i\}$ ($i=1,\dots, N$) define a subspace in
the full Hilbert space of an unconstrained system. A state $\P$ in this
subspace satisfies the constraint equations $\C_i \P=0$. When $\P$ is acted
upon by the observable $A$, one requires that the result $A\P$ remain in
the constrained subspace. The condition for this is
$[A,\C_i]=f_i(\C_1,\ldots,\C_N)$ because then
$$\C_i A\P= -[A,\C_i]\P = -f_i(\C_1,\ldots,\C_N) \P =0 .$$
If one were to weaken the definition of an observable by not requiring that it
commute with the super-Hamiltonian, as is sometimes done\cite{Kuc},
then one must deal with the difficult
problem of operators whose action takes one out of one's
Hilbert space.  This is not an adequate strategy for dealing with the
problem of the frozen formalism; it trades one problem for a harder one.

If the difficulty is not in this definition of an observable, perhaps it
lies in the fact the super-Hamiltonian is a constraint. Constraints are
often a consequence of a symmetry underlying the theory. In the ADM
canonical quantization of gravity, it is well-known that invariance of the
theory under space-time diffeomorphisms makes the super-Hamiltonian
a constraint. In the parametrized canonical formulation of quantum
mechanics, reparametrization invariance of the theory makes the
super-Hamiltonian a constraint. In both
cases, the symmetry making the super-Hamiltonian a constraint is a
physically motivated symmetry which is not to be given up lightly.

The problem of the frozen formalism is thus a real one, at least in so far
as it reflects a weakness in our understanding. It does not however prevent
one from just using the familiar machinery of quantum mechanics. For this
reason, it is most often consigned to the limbo of ``peculiarities of the
quantum formalism,'' and is either dismissed as a problem in semantics or
simply not addressed.

There is without doubt a semantic component to the
problem. In common usage, the word ``observable''
has the connotation ``something which can be observed.'' In ordinary
quantum mechanics, it is defined as a self-adjoint operator with complete
spectrum. In parametrized and constrained quantum mechanics, it is
defined as an operator, not necessarily self-adjoint, which commutes with
all of the constraints. The task is to distinguish these meanings. In so
doing, we shall find that the problem of the frozen formalism is more
subtle than confusing one word with three meanings. It will hinge on how we
describe physical measurements in the mathematical formalism of quantum
mechanics. I will give an explication of the problem by way of a few
examples. These will show that there is no problem with working in the
frozen formalism: there are both constant observables and dynamics within
the wavefunction of the universe.

The essential property of an observable in both its mathematical
definitions is that it has an associated (complete) collection of
eigenstates with corresponding eigenvalues. The significance of this is
that states can be characterized by the eigenvalues of a collection of
commuting observables. The eigenvalues are the quantum numbers of the
state. In the parametrized formalism, these eigenvalues characterize the
state throughout its entire evolution. This is why they are constants of
the motion. If an operator does not commute with the super-Hamiltonian
constraint, its eigenstates are not in the constrained Hilbert space and
are then of no use for representing states in the constrained Hilbert
space.

Because the eigenstates of observables are assumed to be complete, one may
represent states as superpositions of eigenstates. The coefficients in the
superposition will be constant. It is not necessary to know the observables
of which the full state is the eigenstate, though they can be constructed
if it is desired.

To firmly establish this perspective on observables, consider the
para\-metrized free particle with the super-Hamiltonian
\beq
\sH=p_0 +p_1^2.
\eeq
Physical states $|\P\r$ are those which satisfy the super-Hamiltonian
constraint
\beq
\sH\P=0.
\eeq
The operator $p_1$ commutes with the super-Hamiltonian and is an
observable.  Its associated eigenstates may be labelled by the eigenvalue
$k$, where
$$p_1|k\r_1= k|k\r_1,$$
and, in the coordinate representation (assuming the canonical commutation
relations $[q_0,p_0]=i$, $[q_1,p_1]=i$), they are
\beq
\l q_1,q_0|k \r_1= {1\over (2\pi)^{1/2}}e^{ikq_1-ik^2 q_0}.
\eeq
The operator $q_1$ does not commute with the super-Hamiltonian and is not
an observable.  In particular the state $q_1|k\r_1$ does not satisfy the
super-Hamiltonian constraint.

An operator closely related to $q_1$ which
is an observable is
\beq
\label{pos}
q_{1t}=e^{-ip_1^2 (q_0-t)}q_1 e^{ip_1^2 (q_0-t)}=q_1-2p_1(q_0-t).
\eeq
This is the observable which is equal to $q_1$ at time $q_0=t$.  It is
one of Rovelli's ``evolving constants of the motion''\cite{Rov}.  Its
eigenstates are characterized by
$$
q_{1t}|x\r_1 =x|x\r_1.
$$
In the coordinate representation, this is
\beq
\l q_1,q_0| x\r_1 = (4\pi i(q_0-t))^{-1/2} e^{i(q_1-x)^2/4(q_0-t)}.
\eeq
This may be recognized as the Green's function for the free particle,
which reduces to $\delta(q_1-x)$ as $q_0\rightarrow t$.  (The
states are normalized using the usual inner product with respect to
$q_1$, but this won't be discussed here.)

A Gaussian superposition of momentum eigenstates can be formed by
\beq
\label{gauss}
|g;\obar k, a\r_1 =(\pi a/2)^{-1/4} \int dk e^{-(k-\obar k)^2/a} |k \r_1.
\eeq
This has the coordinate representation
$$
\l q_1,q_0|g;\obar k, a\r_1 =( 2\pi a)^{-1/4} (iq_0 +1/a)^{-1/2}
\exp(-\obar k^2 / a)
\exp({i(q_1-2i\obar k/a)^2 \over 4( q_0-i/a)}).
$$
An observable of which this state is an eigenstate is found to  be
\beq
\label{Gobs}
G=q_1-2p_1(q_0-i/a),
\eeq
and the state has eigenvalue $2i\obar k/a$.  Note that $G$ is not
self-adjoint in the usual inner product and its eigenvalue is not real.
One expects that this means that it is not physically observable,
but to confirm this requires a discussion of measurement.

Measurement theory in the foundation of quantum mechanics has been
discussed exhaustively over the past sixty years.  To put the
use of observables as self-adjoint operators in context, it is necessary
to reiterate the litany.  I want to emphasize the central role of
projection
operators in the conventional approach.  In contrast, I want to draw
attention to an argument from a new perspective
compelling the use of a post-Everett description
of measurement in which both system and observing apparatus appear
explicitly.

In ordinary quantum mechanics, observables as self-adjoint operators play a
central role, again through their eigenstates. The conventional
description of measurement is the following:  If one intends to measure a
particular observable, one decomposes the state of the system into a
superposition of eigenstates of that observable. The eigenvalues of these
eigenstates of the observable are interpreted as the possible outcomes of
the measurement. Since the observables are self-adjoint, the eigenvalues,
and hence the outcomes of measurement, are necessarily real. The
probabilities for each of the outcomes are given by the square-modulus of the
coefficients in the superposition. When the measurement is complete, the
state of the system is in an eigenstate of the observable.

This procedure is so ingrained in our understanding of quantum mechanics
that one easily forgets that it is a theoretical construct and not the
measurement process itself. The procedure is primarily based on two {\it
assumptions}\cite{Dir}: 1) measurement of a state gives a particular result
with certainty if and only if the system is in an eigenstate of the
observable being measured, and the result is the eigenvalue of that
eigenstate; 2) from ``physical continuity,'' after a measurement is made,
if that measurement is immediately repeated, the same outcome must be
obtained with certainty, and, hence, by 1), the measurement must put the
system into an eigenstate of the observable. These two assumptions
characterize measurements, distinguishing them from other interactions, and
are thus the fundamental tie between the physically observed and the
mathematically observable, between measurement outcomes and eigenstates of
operators. Few would doubt the validity of the assumptions. I do not claim
that the procedure does not work, but rather that it works too well.

Let us call this description of measurement ``the projection procedure,'' as
one pro\-jects the initial state onto the eigenstates of the observable
being measured.  This projection procedure neatly summarizes the
results of measurement, but does so at the cost of neglecting a
description of the process by which the measurement is made.
It is as if an external agent is able to effect a measurement on the
system without need of introducing any apparatus:  suddenly, the measurement
is done.   The description is wholly isolated.  Only the system is
present, and the measurement has direct access to its state.
Unfortunately, we do not share this luxury of direct access to states.  By
necessity, we must always employ intermediaries to investigate the state of a
system.

A question that we are accustomed to ask in quantum mechanics is
\begin{quote}
``What is the probability density that the momentum of particle-1
in state $|\P\r_1$ is $k$?''
\end{quote}
Suppose the state $|\P\r_1$ is the gaussian superposition of momentum
eigenstates (\ref{gauss}) in the example above. The question inquires
directly about the state of particle-1, and, in the projection procedure,
the question is meaningful and has the familiar answer
$(\pi a/2)^{-1/2} e^{-2(k-\obar k)^2/a}$.
This is not however an entirely sensible question in the context
of a system described by a super-Hamiltonian constraint. To verify the
answer, we must conduct an experiment. The state solving the
super-Hamiltonian constraint is the wavefunction of the universe and
contains, along with everything else, all measurements and their outcomes.
In fact, no measurements were ever made. The question has no truth value
because its answer can be neither confirmed nor denied.

To address the question, additional subsystems must be introduced which
interact with particle-1 to produce the measurement. For the purposes of
theory, these additional subsystems may be hypothetical, as we need not do
every experiment we contemplate, but we must augment the hypothesized
super-Hamiltonian as if the experiment were to be performed. In the event
that it is, we can then expect to confirm or deny our theoretical result.
This treatment of the super-Hamiltonian carries an important resonance
with Bohr's insistence that reality
is determined by the full experimental arrangement\cite{Boh}: the choice
of experiments determines the super-Hamiltonian; the super-Hamiltonian
(plus initial conditions)
determines the wavefunction of the universe and hence reality.

An essential consequence of this is that, to understand the
measurement process properly, one must model the interaction.
It is not enough to add apparatus
subsystems to the
super-Hamiltonian if one continues to treat measurement as a
black box which spontaneously changes the combined system and apparatus state
from an
uncorrelated to a correlated superposition.  This is essentially still
the projection procedure, albeit without the final selection of a
particular term from the correlated superposition.

Before investigating such a model explicitly, consider the characteristics
it must possess.  Our goal is to understand the relation between observables as
self-adjoint operators and physical measurements.   As the correspondence
between them is made through the assumptions underlying the projection
procedure, we desire a model which is as close to the projection
procedure as possible while being more specific about the details of the
interaction. In particular, we require that a measurement of a chosen
observable return a result which distinguishes between different
eigenstates of the observable and that it have the property that if the
measurement is immediately repeated, the same result will be found with
certainty.  This type of model was discussed by von
Neumann\cite{vNe} and plays an important role
in the Everett
interpretation\cite{DeG}. I will discuss it again to emphasize certain
features.

If one has an isolated state being observed without apparatus,
as in the projection procedure, the only quantity which distinguishes
between eigenstates of an observable are their eigenvalues. This is why a
measurement in the projection procedure must return the eigenvalue of the
eigenstate. In a more general setting, in which the state of one subsystem
interacts with another to perform a measurement, the result need only be
a (non-degenerate) correlation of the states of the observing subsystem
with the
eigenstates of the observable in the observed subsystem. This
correlation allows one to infer the state of one system from the state of
the other.  Since the eigenstates of the observed subsystem are
characterized by their eigenvalues, one may say that the measurement has
returned the eigenvalue, in the sense that the eigenvalue
can be inferred from knowledge of the state of the observing subsystem.
This is however an abstraction: the eigenvalue is not an
extant physical quantity.  The physical result of a measurement is the
correlation of the states of subsystems.

The second criterion---that if the measurement is immediately repeated, the
same result is obtained with certainty---is a requirement that the
measurement be non-destructive\cite{DeG}.  That is, if the observed
subsystem is in
an eigenstate of the observable, this eigenstate
must be preserved after the interaction, so that it may be
measured again and found to give the same result.  This rules out, as
measurements, interactions which correlate the state of the observing
subsystem with the state of the observed system before the interaction
but leave it disturbed after the interaction.  As one might expect, this
restricts the interaction terms that may be classified as measurements in
the projection procedure sense.  This is significant because it reveals
that the projection procedure is an idealization of the process of
measurement.  There are interactions which are considered measurements in
experimental practice that are not measurements in this sense.

A further idealization of the process of measurement in the projection
procedure is that it is instantaneous.  This feature is not retained in
the model system: necessarily all measurements implemented by interaction
require finite duration.  The implications of this regarding observables will
be
discussed below.  I remark here that this is a profound
departure from
the projection procedure in both its Copenhagen and Everett incarnations.
It has been lamented\cite{Sch,vNe,Kuc2} that one of
the most serious failings of the quantum mechanical formalism, especially
from the perspective of relativity, is the fact that measurements take
place at a precise instant of time.  This is where this
begins to change.  Measurements as projections, and as results computed
from expectation values, take place at a precise instant of time.
Measurements as interactions require duration.

In the post-Everett view, where the outcome of a measurement is a correlation
between subsystems, the second criterion is a question of
conditional probability.  One confirms that it is satisfied by using the
Page-Wootters
interpretation\cite{PaW}.  One requires two observing subsystems.
Sequentially, each interacts with the observed subsystem establishing
correlation with the observed subsystem. The question is then posed: given
the result of the first of the measurement, is the probability certain that
the result of the second is the same? The answer is yes, by construction.
When the first observing subsystem interacts with the observed subsystem,
it establishes a correlation which distinguishes the different eigenstates
of the observed subsystem. In the manner in which one handles conditional
probabilities, one discards all the states except for the one whose
correlation reflects the given result of the first measurement. The second
observing subsystem then interacts only with an eigenstate of the
observable, not with a superposition, and establishes a correlation which
is the same as that of the first subsystem. The only thing that could go
wrong would be if the observed subsystem is not still in an eigenstate of
the observable, but the measurement-interaction is chosen so that this
cannot happen.

Consider a model of a measurement of the momentum $p_1$ of particle-1 in the
example above.  We introduce a second free particle, particle-2,
which interacts with particle-1 through the
measurement-interaction (cf. \cite{vNe})
\beq
\label{sHi}
\sH_I=a(q_0)p_1 q_2.
\eeq
Since the interaction couples to the observable, it will preserve the
eigenstates of the observable through the measurement-interaction.
Here, $a(q_0)$ is a smooth function which vanishes outside the interval
$0<q_0<T$ and for which $\int_0^T a(q_0^\prime) dq_0^\prime=1$.
It can be viewed as a phenomenological summary of a more
detailed process by which particle-1 and particle-2 are brought together
to interact.  The full super-Hamiltonian is then
\beq
\sH=p_0 +p_1^2 +p_2^2 +a(q_0)p_1 q_2.
\eeq

This problem can be exactly solved, using for example canonical
trans\-for\-ma\-tions \cite{And} (cf. also \cite{Kuc0}).  Define
\beq
A(q_0)=\left\{ \begin{array}{ll}
                 0 & q_0<0 \\
				 \int_{0}^{q_0} a(q_0^\prime) dq_0^\prime& 0\le q_0 \le T .\\
				 1 & q_0>T
				 \end{array}  \right.
\eeq
The super-Hamiltonian $\sH$ with the interaction
term is related to the super-Hamil\-ton\-ian $\sH_0=p_0+p_1^2+p_2^2$
without interaction term
by a time-dependent canonical transformation $C_{q_0}$,
$$ \sH= C_{q_0}\sH_0 C_{q_0}^{-1},$$
where
\beq
C_{q_0}=e^{-i p_1^2\int_{-\infty}^{q_0} A^2(q_0^\prime) dq_0^\prime}
 e^{-iA(q_0) p_1 q_2}
e^{2ip_1 p_2 \int_{-\infty}^{q_0} A(q_0^\prime) dq_0^\prime}.
\eeq
The solutions $|\P\r$ of $\sH$ are given in terms of those $|\P_0\r$ of
$\sH_0$ by
$$|\P\r=C_{q_0}|\P_0\r. $$
Assume that particle-1 is initially in
an eigenstate of momentum $p_1$, $|k\r_1$, and that particle-2 is in an
eigenstate of momentum $p_2$, $|k_2\r_2$, so that
$$|\P_0\r=|k\r_1 |k_2\r_2.$$
The coordinate representation
of the solution $|\P\r$  is
\begin{eqnarray}
\label{state}
\l q_1,q_2,q_0|\P\r &=& \l q_1,q_2,q_0|C_{q_0}|k\r_1 |k_2\r_2 \\
&=&{1\over 2\pi}\exp(ik q_1 +i (k_2-A(q_0) k) q_2-i(k^2 +k_2^2) q_0
 \nonumber \\
&&\hspace{1cm}+ i2k k_2 \int_{-\infty}^{q_0} A(q_0^\prime) dq_0^\prime
-i k^2\int_{-\infty}^{q_0} A^2(q_0^\prime) dq_0^\prime). \nonumber
\end{eqnarray}
The state evolves smoothly from $|k\r_1 |k_2\r_2$ before $q_0=0$ to
$e^{i\phi(k,k_2)}|k\r_1 |k_2-k\r_2$ after $q_0=T$.
A phase
$\phi(k,k_2)$  arises in the evolution and, explicitly,
$$
\phi(k,k_2)=i2k k_2(c_1-T) -ik^2(c_2-T),
$$
where
$$c_1=\int_0^T A(q_0^\prime) dq_0^\prime$$
and
$$c_2=\int_0^T A^2(q_0^\prime) dq_0^\prime.$$
The state of particle-2 is correlated with that of particle-1 after the
evolution, and the eigenstate of particle-1 has not been disturbed.  A
measurement has been performed.

If particle-1 were initially in the Gaussian
superposition of momentum-eigenstates (\ref{gauss}), the measurement would
have produced the smooth transition to a superposition of correlated states
\begin{eqnarray}
\biggl( (\pi a/2)^{-1/4} \int dk e^{-(k-\obar k)^2/a} |k \r_1 \biggr)
|k_2\r_2 &\longrightarrow & \\
&& \hspace{-2.5cm} (\pi a/2)^{-1/4} \int dk e^{-(k-\obar k)^2/a}
e^{i\phi(k,k_2)}|k \r_1 |k_2-k\r_2. \nonumber
\end{eqnarray}
Suppose one introduces a second observer, particle-3, in the same initial
state $|k_2\r_3$ as particle-2, and couples it to particle-1 for an
interval after $q_0=T$ through a term analogous to (\ref{sHi}).
Given that the result of the first measurement is
$k'$, i.e. the correlation $|k'\r_1 |k_2-k'\r_2$, the result of the second
measurement will be the correlated state $|k'\r_1 |k_2-k'\r_2 |k_2-k'\r_3$.
The same measurement result is obtained, as required.

Let us now consider the role of observables.  In the absence of the
interaction term (\ref{sHi}), both $p_1$ and $p_2$ commute with the
super-Hamiltonian $\sH_0$ and are observables.  In the presence of the
interaction, $p_2$ is no longer an observable.  This is consistent with the
fact
that the initial state of particle-2, which is characterized by its eigenvalue
with respect to $p_2$, changes during the interaction.   Even though $p_2$
is not an observable, a modification gives an observable
\beq
\tilde p_2=C_{q_0}p_2 C_{q_0}^{-1}=p_2 +A(q_0) p_1.
\eeq
The full quantum wavefunction (\ref{state}) over the whole history of the
universe has the
eigenvalue $k_2$ for $\tilde p_2$ and the eigenvalue $k$
for the observable $p_1$.  These eigenvalues label the state,
and they are constants throughout the evolution of the state.
Nevertheless a measurement has been made.  There is no loss of dynamics
because one has chosen to work in the frozen formalism.

A closer examination of the relation between observables and dynamics
will be illuminating.
Note that $\tilde p_2$ agrees with $p_2$ when $q_0<0$. For this restricted
portion of the universe, $p_2$ is an observable in the sense that it
commutes with the super-Hamiltonian, and it can be used to label states in
this region. This suggests that it is useful to distinguish between a
restricted observable which commutes with the super-Hamiltonian in some
region and a global observable which commutes with the super-Hamiltonian
everywhere.

As participants in the universe, we do not of course know the full
super-Hamiltonian which describes it. There will be measurements made in
the future which we cannot anticipate now. Since we only discover the
details of the super-Hamiltonian of the universe as we go along, we cannot
know the global observables which commute with the super-Hamiltonian of our
universe. When we say that the states of subsystems we observe are in
eigenstates of some observables, they are in eigenstates of restricted
observables. For some period of time, those observables commute with the
super-Hamiltonian of the universe, and their eigenstates are unchanging
with respect to eigenstates of other observables that also commute with the
super-Hamiltonian.

To elaborate on this further, consider the observable $p_1$ which is being
measured. In the example here, it is both an observable in the sense that
it commutes with the super-Hamiltonian and in the sense that a correlation
with its eigenstates is established during the measurement-interaction with
particle-2. I want to emphasize that it is not necessary that $p_1$ commute
with the super-Hamiltonian for all $q_0$, so long as it does so in the
neighborhood of the period of measurement.

Suppose one considers the measurement of $p_1$ when the state of particle-1
at $q_0=0$ is the gaussian superposition (\ref{gauss}). One could add a
$q_1$-dependent term to the super-Hamiltonian which evolves some initial
state of particle-1 into the gaussian superposition and turns off before
$q_0=0$, when the measurement begins. Or, one could add such a term some
time after $q_0=T$ when the measurement is complete, and the final state of
particle-1 in each correlated state of the superposition would evolve away
from a momentum eigenstate. In each case, the momentum of particle-1 in the
gaussian superposition state at $q_0=0$ would still be measured, but $p_1$
would only be a restricted observable. It would not commute with the
super-Hamiltonian if there were $q_1$-dependent terms present.
Not being a global observable means that the eigenvalue of $p_1$ could not
be used as a quantum number for the wavefunction of the universe, but this
is not a serious loss. If one's primary concern is with predictions of
the outcomes of measurement, restricted
observables are more relevant than global ones.

The nature of observables can be still more closely investigated.
At each instant $q_0=t$, the state of particle-2 is instantaneously an
eigenstate of the self-adjoint operator $p_2$ with eigenvalue $k_2
-A(t)k$. In the ordinary
quantum mechanical sense, $p_2$ is an observable.  One can compute
expectation values of it at
any time $q_0$,
and one thinks of these as predictions of the outcomes of possible
measurements.  Now, $p_2$
is not a global observable, and it doesn't commute with
$\sH$ at $q_0=t$ when $0<t<T$, so it isn't always a restricted observable.
Nevertheless, just as $q_1$ at time $q_0=t$, in the first example, was made
into a global observable above by evolving it with the Hamiltonian, $p_2$
can be made a global observable by applying the canonical
transformation $C_{q_0}C_t^{-1}$. The observable is
\beq
p_{2t}=C_{q_0}C_t^{-1}p_2 C_t C_{q_0}^{-1}=p_2+A(q_0)p_1-A(t)p_1 .
\eeq
This gives a family of observables $p_{2t}$ which reduce to the operator
$p_2$ at time
$q_0=t$ of which particle-2 is
instantaneously an eigenstate.  As the state of the system evolves
through the measurement, the eigenstate of particle-2 changes at
each instant as the observable of which it is the eigenstate changes.
In ordinary quantum mechanics, when one
speaks of the self-adjoint
operator $p_2$ as an observable, one is referring to $p_{2t}$.

Incidentally, this answers
Kuchar's criticism that the Page-Wootters conditional probability
interpretation does
not give the correct answer for prop\-a\-gators\cite{Kuc}.  The observables for
the
position at two distinct instants of time are different, as given by
(\ref{pos}).  If, at time $q_0=T$, one wants to
predict the probability of finding the particle at some location at a
later instant $q_0=T'$, one
must compute the conditional probability that the particle is in an
eigenstate of $q_{1T'}$.  If one uses $q_{1T}$ as the position observable for
all time, the particle will not appear to move, as Kuchar rightly argues.

It is generally true that an operator at
an instant of time can be promoted into a global observable, and hence
one has a family of observables parametrized by the time.  These are
Rovelli's evolving constants of the motion\cite{Rov}.  As these
observables change, the eigenstates associated with them change as well.
This change embodies the evolution of states.

One may ask whether these observables are all physically measurable.  That
is, can one introduce an observing subsystem that will correlate with the
momentum of particle-2 at time $q_0=t$ for $0<t<T$?  My answer is no.
While one may formally calculate expectation values for the momentum
$p_{2t}$ at these times, these calculations do not refer to the
results of any physical
experiment that can be done, in the projection procedure sense.  There
are two related difficulties.  First, all physical measurements require finite
duration in order to establish correlations between the observing and the
observed subsystems.  This is itself a subject requiring further
elaboration, but for the moment suffice it to say that, since the
eigenvalue of the operator $p_2$ is changing,
an attempted measurement can at best measure an averaged
value and not the specific momentum at time $q_0=t$.  Moreover, one expects
that no coupling exists which will leave the
changing value of $p_2$ undisturbed, so that the measurement of $p_1$ is
unaffected.  Secondly, because
$p_2$ is dynamically changing, it is impossible to arrange that a second
measurement will find the same result with certainty.  One can
couple to the observable which corresponds to the instantaneous momentum
eigenstate of particle-2 at time $q_0=t$, but as it will obtain an
average result over a different interval than the first measurement, the
results will in general be different.  This would not then be a measurement in
the projection procedure sense.

Thus, only restricted observables can be physically measured in the
projection procedure sense. One is led to the conclusion that the
assumptions about the nature of measurement that lie at the foundation of
the projection procedure are too idealized. By postulating instantaneous
non-disruptive measurement, they both exclude physically relevant
measurement-interactions and allow computations for the outcomes of
experiments that cannot be realized. It is evident that further work on
measurement theory outside the projection procedure framework is
necessary.

To close one final loose-end, consider whether the non-self-adjoint
observable $G$ (\ref{Gobs}) can be physically observed. Mathematically, the
answer would seem to be yes: one uses a coupling analogous to (\ref{sHi})
with $p_1$ replaced by $G$. This would establish a correlation between the
state of particle-2 and the $G$-eigenstate of particle-1. There is however
a difficulty. Since $G$ is a complex operator, it is not evident that there
exists a physical device which can realize the proposed coupling. This
serves to emphasize a very important point. In the laboratory, we are
restricted to a handful of possible interactions. One must bear in mind
that these are the building blocks from which we must ultimately build our
super-Hamiltonian.

The following picture of dynamics in the frozen formalism can be assembled
from the foregoing discussion. The full quantum state representing the
``wavefunction of the universe'' is fixed once the initial conditions and
the super-Hamiltonian are given. This includes all measurements that will
be made during the course of the universe. Dynamical evolution is a process
that takes place in the form of changes in the decomposition of the full
state into subsystem eigenstates. The wavefunction of the universe need not
be expressed as a product state of eigenstates of its global observables.
It may of course be represented as a superposition of such eigenstates.
More generally it may be represented in terms of eigenstates of operators
which are observables only in restricted regions of the universe, or in
terms of eigenstates of families of global observables parametrized by the
time. When the wavefunction of the universe is expressed in such a fashion,
one finds that as the collection of observables used to decompose the state
change, the superposition of eigenstates change. This is what gives us the
impression of dynamical evolution: it is the changing collection of
correlations amongst the eigenstates of restricted observables that
constitutes what we observe.

The self-adjoint operators that we speak of in ordinary quantum mechanics
as observables are members of families of global observables parametrized
by the time. Because any measurement made through interactions requires
finite duration to establish correlations between the observing and the
observed subsystems, only restricted observables which commute with the
super-Hamiltonian through the period of measurement are physically
measurable, in the projection procedure sense. In particular, this means
that one can compute expectation values for many self-adjoint operators
which do not refer to the outcomes of physically realizable experiments. If
one is interested in physics, care must be taken with the use of
expectation values. More importantly, one must appreciate that the
projection procedure, which so strongly colors our perception of quantum
mechanics, overly idealizes measurement as instantaneous and
non-destructive. Recognizing that a proper description of measurements
within the quantum formalism requires interactions of finite duration is a
first step towards resolving the long-standing conflict over the role of
time in quantum mechanics and relativity.
\vskip 1cm

I would like to thank A. Albrecht and C.J. Isham for discussions
improving the presentation of this work.

\bibliographystyle{plain}

\end{document}